\documentclass[sigconf]{acmart}

\AtBeginDocument{%
  }

\begin{document}

\title[Rs4rs: A Recommender System for Recommendation System Research]{Rs4rs: Semantically Find Recent Publications from Top Recommendation System-Related Venues}

\author{Tri Kurniawan Wijaya}
\affiliation{
  \institution{Huawei Ireland Research Centre}
  \city{Dublin}
  \country{Ireland}}
\email{tri.kurniawan.wijaya@huawei.com}

\author{Edoardo D'Amico}
\affiliation{
  \institution{Huawei Ireland Research Centre}
  \city{Dublin}
  \country{Ireland}}
\email{edoardo.damico@huawei-partners.com}

\author{Gabor Fodor}
\affiliation{
  \institution{Huawei Ireland Research Centre}
  \city{Dublin}
  \country{Ireland}}
\email{gabor.fodor@huawei-partners.com}

\author{Manuel V. Loureiro}
\affiliation{
  \institution{Huawei Ireland Research Centre}
  \city{Dublin}
  \country{Ireland}}
\email{manuel.loureiro@huawei.com}


\begin{abstract}
Rs4rs is a web application designed to perform semantic search on recent papers from top conferences and journals related to Recommender Systems. 
Current scholarly search engine tools like Google Scholar, Semantic Scholar, and ResearchGate often yield broad results that fail to target the most relevant high-quality publications. 
Moreover, manually visiting individual conference and journal websites is a time-consuming process that primarily supports only syntactic searches. 
Rs4rs addresses these issues by providing a user-friendly platform where researchers can input their topic of interest and receive a list of recent, relevant papers from top Recommender Systems venues. 
Utilizing semantic search techniques, Rs4rs ensures that the search results are not only precise and relevant but also comprehensive, capturing papers regardless of variations in wording. 
This tool significantly enhances research efficiency and accuracy, thereby benefitting the research community and public by facilitating access to high-quality, pertinent academic resources in the field of Recommender Systems. 
Rs4rs is available at \url{https://rs4rs.com}.

\end{abstract}




\maketitle

\section{Introduction}

In the rapidly evolving field of Recommender Systems, keeping up with the latest research is essential but 
challenging~\cite{bir-2024}. 
Current scholarly search engines (such as 
Google Scholar~\cite{google-scholar}, 
Semantic Scholar~\cite{semantic-scholar}, 
ResearchGate~\cite{research-gate}, 
Connectedpapers~\cite{connected-papers},
SciSpace~\cite{scispace}, and 
Scite~\cite{scite})
often provide overly broad search results that may overlook the most relevant and high-quality publications in Recommender Systems. 
Additionally, the process of manually navigating through various Recommender System-related conferences and journal websites to find pertinent research is not only time-consuming but also typically limited to basic keyword searches. 
This approach often fails to capture the context of the research, leading to less effective literature retrieval. 
This inefficiency can significantly hinder research progress, especially in the fast-paced field of Recommender Systems. 

To address these challenges, we present \emph{Rs4rs}, a web application designed for performing semantic searches on recent papers from top conferences and journals related to Recommender Systems. 
Rs4rs enhances research efficiency and accuracy by offering a user-friendly platform where researchers can input their topics of interest, filter results based on specific conferences, and receive a curated list of recent, pertinent papers. Additionally, Rs4rs allows users to share search results via a link and provides direct redirection to the full text of the papers.

Despite recent progress in bibliographic and scholarly article search and 
retrieval~\cite{bottoni2023graph, rybinski2023sciharvester, roy2024gear, nilles2024conversational, lahav2022search, bir-2024}, Rs4rs offers a unique contribution, which lies in its specialized focus on Recommender Systems. 
Rs4rs enhances the search experience by retrieving papers exclusively 
from top-tier conferences and journals related to Recommender Systems, 
ensuring that users access high-quality and relevant research. 
Additionally, its use of semantic search techniques allows for more accurate and context-aware retrieval, 
providing researchers with precise and comprehensive results 
tailored to their specific needs in the Recommender Systems domain.
In short, Rs4rs aims to accelerate the literature review process, enable researchers to quickly identify emerging trends and relevant work, and ultimately contribute to faster advancement of the field.

\section{System Design}

\begin{figure*}
  \centering
  \includegraphics[width=0.8\linewidth]{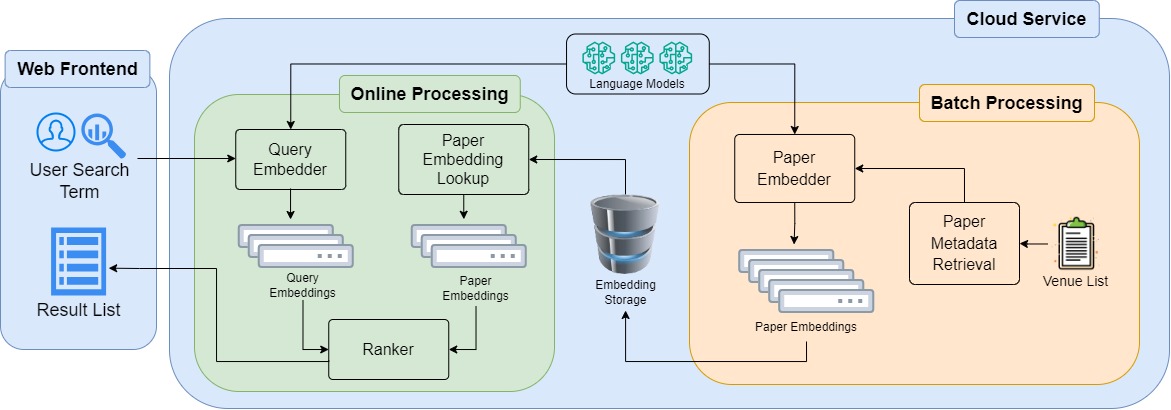}
  \caption{Rs4rs Architecture - Online Processing and Batch Processing Components.}
  \Description{System Architecture of Rs4rs}
  \label{fig:arch}
\end{figure*}



\subsection{System Requirements}
Rs4rs is  designed with a set of requirements, each carefully chosen to address the unique challenges faced by researchers in the field of Recommender Systems. 

First, the system should exclusively index papers from \emph{top conferences and journals related to Recommender Systems}. This focus ensures that users are presented with high-quality, relevant papers, saving time and effort in filtering through non-relevant literature.

Second, it should prioritize indexing \emph{recent papers}, with an initial focus on papers from the last few (two) years. 
This is because recent papers are more likely to reflect the latest advancements and trends in the field, making the search results more relevant to current research needs.
The capability to expand the database to include older papers should be available in the future. 

Third, it should use \emph{batch processing} for the initial indexing and periodic updates of the literature database and the \emph{online processing} for showing relevant papers based on user queries.


\subsection{System Architecture}
Figure~\ref{fig:arch} shows the architecure of Rs4rs.

\noindent
\textbf{Batch Processing vs. Online Processing.}
Rs4rs' \emph{batch processing} retrieves a comprehensive list of papers and pre-compute their embeddings
to significantly reduces the computational load during \emph{online processing}, 
allowing Rs4rs to quickly respond to user queries and rank relevant papers in real-time. 

\noindent
\textbf{Accepts Multiple Encodings.}
%
%
%
Rs4rs is designed to be flexible in 
accommodating multiple language models to potentially 
improve ranking quality through model ensembling. 
To be cost-effective, Rs4rs uses the top two SBERT~\cite{reimers-2019-sentence-bert,sbert-pretrained} models, 
\emph{all-mpnet-base-v2} and \emph{multi-qa-mpnet-base-dot-v1}, instead of a large language model. 
These small, fine-tuned models specialize in semantic similarity. 
We average the cosine similarity of the query and paper embeddings from both models. 
The higher the similarity to the user query, the higher the paper ranks. 
We plan to systematically measure the quality of various language model combinations in the future.

\noindent
\textbf{Data Sources and Retrieval}
The list of top Recommender System venues is sourced from \emph{recsys.info}~\cite{recsys-info}, 
ensuring that Rs4rs focuses on the top conferences and journals in the field of 
Recommender Systems. 
For each venue, Rs4rs retrieves paper titles from \emph{DBLP}~\cite{dblp} 
and abstracts from \emph{Semantic Scholar API}~\cite{semantic-scholar-api}.


\section{Key Features}
\noindent
\textbf{High-Quality Source Selection.}
Rs4rs guarantees that search results are drawn from a highly selective pool of sources. 
By restricting the database to papers from top-tier (A* and A-ranked~\cite{recsys-info}) 
conferences and journals specifically related to Recommender Systems, 
it increases the likelihood for its search result to also be of high-quality. 

\noindent
\textbf{Semantic Search.}
Rs4rs ensures that users can find relevant papers even if they make typographical errors or use synonyms different from those in the actual papers. This feature makes the search process intuitive and user-friendly, allowing researchers to focus on discovering valuable content without worrying about precise keyword matching.

\noindent
\textbf{Customized Filtering Options.}
Rs4rs empowers users to select and filter results based on specific conferences or journals, giving them the flexibility to narrow down their search to the most relevant and impactful publications they prefer.





\section{Use Cases and Applications}
Rs4rs is designed to cater to the needs of various users within the Recommender Systems community. 
Typical beneficiary profiles includes, but not limited to, 
1) students who are in the early stages of their Recommender Systems research, or
2) experienced researchers who wish to explore new domains within Recommender Systems or keep abreast of developments in adjacent areas, or 
3) authors looking to update their survey papers and ensure their reviews include the most recent advancements.

Typical workflow with Rs4rs is as follows:
\begin{itemize}
\item Users start by entering a search term or topic of interest, such as "LLM for Recommender Systems."
\item The resulting publications are displayed with sentences from the abstracts included. 
\item Users can expand or hide the abstract sections based on their preference using the drop-down menu provided.
\item Users can click on the year column headers to sort the results, helping them track research developments over time.
\item For more experienced users, there is an option to filter publications by entering their preferred venues in the filter box, refining the search to meet their specific requirements.
\end{itemize}

 


\section{Conclusion}
Rs4rs is a publicly-available web application for semantic search across recent papers from top-tier Recommender Systems conferences and journals. 
The key contribution of Rs4rs is in its ability to streamline the research process by providing highly relevant and quality-filtered academic papers, 
thereby relieving researchers from having to sift through less pertinent or lower-quality content. 

Looking forward, there are several promising directions for future work. 
Its semantic search could still be refined to enhance the result relevance. 
An email subscription services or RSS-like feeds could be added to 
allow users to receive notifications about new publications that match their research interests. 
Additionally, Rs4rs can be expanded to support more sophisticated research activities, 
such as meta-analyses, trend detection, and mapping research trajectories within the Recommender Systems domain. 
%
As we continue to enhance Rs4rs, we invite feedback to ensure the tool evolves in ways that best support the needs and aspirations of the Recommender Systems community. 
\bibliographystyle{ACM-Reference-Format}
\bibliography{main}


\begin{thebibliography}{17}


\ifx \showCODEN    \undefined \def \showCODEN     #1{\unskip}     \fi
\ifx \showDOI      \undefined \def \showDOI       #1{#1}\fi
\ifx \showISBNx    \undefined \def \showISBNx     #1{\unskip}     \fi
\ifx \showISBNxiii \undefined \def \showISBNxiii  #1{\unskip}     \fi
\ifx \showISSN     \undefined \def \showISSN      #1{\unskip}     \fi
\ifx \showLCCN     \undefined \def \showLCCN      #1{\unskip}     \fi
\ifx \shownote     \undefined \def \shownote      #1{#1}          \fi
\ifx \showarticletitle \undefined \def \showarticletitle #1{#1}   \fi
\ifx \showURL      \undefined \def \showURL       {\relax}        \fi
\providecommand\bibfield[2]{#2}
\providecommand\bibinfo[2]{#2}
\providecommand\natexlab[1]{#1}
\providecommand\showeprint[2][]{arXiv:#2}

\bibitem[Bottoni et~al\mbox{.}(2023)]%
        {bottoni2023graph}
\bibfield{author}{\bibinfo{person}{Simone Bottoni}, \bibinfo{person}{Alberto Trombetta}, \bibinfo{person}{Flavio Bertini}, \bibinfo{person}{Danilo Montesi}, \bibinfo{person}{Francesca Bonin}, \bibinfo{person}{Alessandra Pascale}, \bibinfo{person}{Martin Gleize}, {and} \bibinfo{person}{Pierpaolo Tommasi}.} \bibinfo{year}{2023}\natexlab{}.
\newblock \showarticletitle{Graph-based Tool for Exploring PubMed Knowledge Base}. In \bibinfo{booktitle}{\emph{2023 IEEE 39th International Conference on Data Engineering (ICDE)}}. IEEE, \bibinfo{pages}{3611--3614}.
\newblock


\bibitem[ConnectedPapers(2024)]%
        {connected-papers}
\bibfield{author}{\bibinfo{person}{ConnectedPapers}.} \bibinfo{year}{2024}\natexlab{}.
\newblock \bibinfo{title}{Find and explore academic papers}.
\newblock \bibinfo{howpublished}{\url{https://www.connectedpapers.com}}.
\newblock
\newblock
\shownote{Accessed: 2024-07-23}.


\bibitem[Dagstuhl(2024)]%
        {dblp}
\bibfield{author}{\bibinfo{person}{Schloss Dagstuhl}.} \bibinfo{year}{2024}\natexlab{}.
\newblock \bibinfo{title}{dblp: computer science bibliography}.
\newblock \bibinfo{howpublished}{\url{https://dblp.org/}}.
\newblock
\newblock
\shownote{Accessed: 2024-07-23}.


\bibitem[for AI(2024a)]%
        {semantic-scholar}
\bibfield{author}{\bibinfo{person}{Allen~Institute for AI}.} \bibinfo{year}{2024}\natexlab{a}.
\newblock \bibinfo{title}{Semantic Scholar}.
\newblock \bibinfo{howpublished}{\url{https://www.semanticscholar.org/}}.
\newblock
\newblock
\shownote{Accessed: 2024-07-23}.


\bibitem[for AI(2024b)]%
        {semantic-scholar-api}
\bibfield{author}{\bibinfo{person}{Allen~Institute for AI}.} \bibinfo{year}{2024}\natexlab{b}.
\newblock \bibinfo{title}{Semantic Scholar Academic Graph API}.
\newblock \bibinfo{howpublished}{\url{https://www.semanticscholar.org/product/api}}.
\newblock
\newblock
\shownote{Accessed: 2024-07-23}.


\bibitem[Frommholz et~al\mbox{.}(2024)]%
        {bir-2024}
\bibfield{author}{\bibinfo{person}{Ingo Frommholz}, \bibinfo{person}{Philipp Mayr}, \bibinfo{person}{Guillaume Cabanac}, {and} \bibinfo{person}{Suzan Verberne}.} \bibinfo{year}{2024}\natexlab{}.
\newblock \showarticletitle{Bibliometric-Enhanced Information Retrieval: 14th International BIR Workshop (BIR 2024)}. In \bibinfo{booktitle}{\emph{Advances in Information Retrieval: 46th European Conference on Information Retrieval, ECIR 2024, Glasgow, UK, March 24–28, 2024, Proceedings, Part V}} (Glasgow, United Kingdom). \bibinfo{publisher}{Springer-Verlag}, \bibinfo{address}{Berlin, Heidelberg}, \bibinfo{pages}{442–446}.
\newblock
\showISBNx{978-3-031-56068-2}
\urldef\tempurl%
\url{https://doi.org/10.1007/978-3-031-56069-9_61}
\showDOI{\tempurl}


\bibitem[Google(2024)]%
        {google-scholar}
\bibfield{author}{\bibinfo{person}{Google}.} \bibinfo{year}{2024}\natexlab{}.
\newblock \bibinfo{title}{Google Scholar}.
\newblock \bibinfo{howpublished}{\url{https://scholar.google.com/}}.
\newblock
\newblock
\shownote{Accessed: 2024-07-24}.


\bibitem[Jain et~al\mbox{.}(2024)]%
        {scispace}
\bibfield{author}{\bibinfo{person}{Siddhant Jain}, \bibinfo{person}{Asheesh Kumar}, \bibinfo{person}{Trinita Roy}, \bibinfo{person}{Kartik Shinde}, \bibinfo{person}{Goutham Vignesh}, {and} \bibinfo{person}{Rohan Tondulkar}.} \bibinfo{year}{2024}\natexlab{}.
\newblock \showarticletitle{SciSpace Literature Review: Harnessing AI for Effortless Scientific Discovery}. In \bibinfo{booktitle}{\emph{Advances in Information Retrieval: 46th European Conference on Information Retrieval, ECIR 2024, Glasgow, UK, March 24–28, 2024, Proceedings, Part V}} (Glasgow, United Kingdom). \bibinfo{publisher}{Springer-Verlag}, \bibinfo{address}{Berlin, Heidelberg}, \bibinfo{pages}{256–260}.
\newblock
\showISBNx{978-3-031-56068-2}
\urldef\tempurl%
\url{https://doi.org/10.1007/978-3-031-56069-9_28}
\showDOI{\tempurl}


\bibitem[Lahav et~al\mbox{.}(2022)]%
        {lahav2022search}
\bibfield{author}{\bibinfo{person}{Dan Lahav}, \bibinfo{person}{Jon~Saad Falcon}, \bibinfo{person}{Bailey Kuehl}, \bibinfo{person}{Sophie Johnson}, \bibinfo{person}{Sravanthi Parasa}, \bibinfo{person}{Noam Shomron}, \bibinfo{person}{Duen~Horng Chau}, \bibinfo{person}{Diyi Yang}, \bibinfo{person}{Eric Horvitz}, \bibinfo{person}{Daniel~S Weld}, {et~al\mbox{.}}} \bibinfo{year}{2022}\natexlab{}.
\newblock \showarticletitle{A search engine for discovery of scientific challenges and directions}. In \bibinfo{booktitle}{\emph{Proceedings of the AAAI Conference on Artificial Intelligence}}, Vol.~\bibinfo{volume}{36}. \bibinfo{pages}{11982--11990}.
\newblock


\bibitem[Nilles(2024)]%
        {nilles2024conversational}
\bibfield{author}{\bibinfo{person}{Markus Nilles}.} \bibinfo{year}{2024}\natexlab{}.
\newblock \showarticletitle{Conversational Bibliographic Search}. In \bibinfo{booktitle}{\emph{Proceedings of the 2024 Conference on Human Information Interaction and Retrieval}}. \bibinfo{pages}{445--448}.
\newblock


\bibitem[RecSys.info(2024)]%
        {recsys-info}
\bibfield{author}{\bibinfo{person}{RecSys.info}.} \bibinfo{year}{2024}\natexlab{}.
\newblock \bibinfo{title}{Track Top Recommender System Conferences and Journals}.
\newblock \bibinfo{howpublished}{\url{https://recsys.info}}.
\newblock
\newblock
\shownote{Accessed: 2024-07-23}.


\bibitem[Reimers and Gurevych(2019)]%
        {reimers-2019-sentence-bert}
\bibfield{author}{\bibinfo{person}{Nils Reimers} {and} \bibinfo{person}{Iryna Gurevych}.} \bibinfo{year}{2019}\natexlab{}.
\newblock \showarticletitle{Sentence-BERT: Sentence Embeddings using Siamese BERT-Networks}. In \bibinfo{booktitle}{\emph{Proceedings of the 2019 Conference on Empirical Methods in Natural Language Processing}}. \bibinfo{publisher}{Association for Computational Linguistics}.
\newblock
\urldef\tempurl%
\url{https://arxiv.org/abs/1908.10084}
\showURL{%
\tempurl}


\bibitem[ResearchGate(2024)]%
        {research-gate}
\bibfield{author}{\bibinfo{person}{ResearchGate}.} \bibinfo{year}{2024}\natexlab{}.
\newblock \bibinfo{title}{Find and share research}.
\newblock \bibinfo{howpublished}{\url{https://researchgate.net/}}.
\newblock
\newblock
\shownote{Accessed: 2024-07-24}.


\bibitem[Roy et~al\mbox{.}(2024)]%
        {roy2024gear}
\bibfield{author}{\bibinfo{person}{Kaushik Roy}, \bibinfo{person}{Vedant Khandelwal}, \bibinfo{person}{Valerie Vera}, \bibinfo{person}{Harshul Surana}, \bibinfo{person}{Heather Heckman}, {and} \bibinfo{person}{Amit Sheth}.} \bibinfo{year}{2024}\natexlab{}.
\newblock \showarticletitle{GEAR-Up: Generative AI and External Knowledge-Based Retrieval: Upgrading Scholarly Article Searches for Systematic Reviews}. In \bibinfo{booktitle}{\emph{Proceedings of the AAAI Conference on Artificial Intelligence}}, Vol.~\bibinfo{volume}{38}. \bibinfo{pages}{23823--23825}.
\newblock


\bibitem[Rybinski et~al\mbox{.}(2023)]%
        {rybinski2023sciharvester}
\bibfield{author}{\bibinfo{person}{Maciej Rybinski}, \bibinfo{person}{Stephen Wan}, \bibinfo{person}{Sarvnaz Karimi}, \bibinfo{person}{Cecile Paris}, \bibinfo{person}{Brian Jin}, \bibinfo{person}{Neil Huth}, \bibinfo{person}{Peter Thorburn}, {and} \bibinfo{person}{Dean Holzworth}.} \bibinfo{year}{2023}\natexlab{}.
\newblock \showarticletitle{SciHarvester: searching scientific documents for numerical values}. In \bibinfo{booktitle}{\emph{Proceedings of the 46th International ACM SIGIR Conference on Research and Development in Information Retrieval}}. \bibinfo{pages}{3135--3139}.
\newblock


\bibitem[Scite(2024)]%
        {scite}
\bibfield{author}{\bibinfo{person}{Scite}.} \bibinfo{year}{2024}\natexlab{}.
\newblock \bibinfo{title}{AI for Research}.
\newblock \bibinfo{howpublished}{\url{https://www.scite.ai}}.
\newblock
\newblock
\shownote{Accessed: 2024-07-23}.


\bibitem[UKPLab(2024)]%
        {sbert-pretrained}
\bibfield{author}{\bibinfo{person}{UKPLab}.} \bibinfo{year}{2024}\natexlab{}.
\newblock \bibinfo{title}{SBERT Pretrained Models}.
\newblock \bibinfo{howpublished}{\url{https://sbert.net/docs/sentence_transformer/pretrained_models.html}}.
\newblock
\newblock
\shownote{Accessed: 2024-07-23}.


\end{thebibliography}










\end{document}